\documentclass[twocolumn]{rmaa}
\usepackage{psfig}
\usepackage{psfrag,color}
\usepackage{booktabs}
\usepackage{longtable}
\usepackage{natbib}

\def\uv{uvby-\beta \ }

\title{$\uv$ photoelectric photometry of NGC 7063}

\author{J. H. Pe\~na\altaffilmark{1}, L. Fox Machado\altaffilmark{2} and R. Garrido\altaffilmark{3}}

\altaffiltext{1}{Instituto de Astronom\'{\i}a, Universidad Nacional
Aut\'onoma de M\'exico, M\'exico, Apdo. Postal 70-264,
M\'exico.}\altaffiltext{2}{ Observatorio Astron\'omico Nacional,
Instituto de Astronom\'{\i}a, Universidad Nacional Aut\'onoma de
M\'exico, Ensenada B.C., Apdo. Postal 877, M\'exico}
\altaffiltext{3}{Instituto de Astrof\'{\i}sica de Andaluc\'{\i}a,
Spain}

\resumen{Mediante fotometr\'{\i}a de 75 estrellas en la direcci\'on
de NGC 7063 se ha determinado la membresia de algunas estrellas,
establecido su distancia ($722 \pm 105$ pc), su edad ($\log$ age de
$8.146$) y su enrojecimiento ($E(b-y) = 0.091 \pm 0.039$ mag).}

\abstract{From $uvby$ photometry of 75 stars in the direction of NGC
7063 we were able to determine membership of some stars and fix the
distance ($722 \pm 105$ pc), $\log$ age (of $8.146$) and reddening
($E(b-y) = 0.091 \pm 0.039$ mag) for the cluster.}


\shortauthor{Pe\~na, et al.} \shorttitle{}

\fulladdresses{ \item J. H. Pe\~na: Instituto de Astronom\'{\i}a,
UNAM, Apdo. Postal 70-264, M\'exico, D.F. M\'exico
(jhpena@astroscu.unam.mx).}

\begin{document}

\maketitle

\section{Introduction}

As a continuation of a study of open clusters and the short period
variable stars within, we now present  the observations done on the
cluster NGC 7063. This relatively poorly populated cluster has
remained practically unstudied.

Collinder (1931) presented the initial study of NGC 7063 and
described a cluster with a diameter of 8' x 5' which contains only
10 to 12 stars at a distance of  $2600$ pc. This distance was later
modified by Johnson et al. (1961) who determined a reddening of
$E(B-V) = 0.08$ and a distance of $630$ pc. UBV measurements of 28
stars were reported by Hoag et al. (1961) and Svoupoulus (1962)
found the same reddening and distance, but from spectroscopy he
determined the following values for distance and reddening: $660$ pc
and $E(B-V) = 0.05$. In 1965 Hoag et al. (1965) gave new spectral
classification for five stars. Schneider (1987) presented $\uv$
measurements of 19 stars which were taken from the list of Hoag et
al. (1961) with $B-V \leq 0.4$ and brighter than $12.0$ mag. He
determined a color excess of $E(b-y) = 0.062 \pm 0.007$ (which
corresponds to $E(B-V) = 0.08$) and a distance modulus of $9.01 \pm
0.09$ or, correspondingly, a distance of $635 \pm 30$ pc, in
agreement with the previously determined values.

\section{Observations}

The instrumentation utilized has the advantage that the $uvby$
photometry is acquired simultaneously and the N and W filters that
define H$\beta$ almost simultaneously. These were all taken at the
Observatorio Astron\'omico Nacional, M\'exico. The $1.5$ m telescope
to which a spectrophotometer was attached was utilized at all times.
The observing seasons were carried out in three runs: August 1986
(66 stars, observers: JHP and R. Garrido); October, 1989 (21 stars
on the night of October 28$^{{\rm th}}$, observers: R. Peniche and
JHP) and July, 2006 (17 stars in the night of July 17$^{{\rm th}}$,
observer: JHP). In each season the same criteria for the selection
of the stars was followed: almost all of the brightest stars up to a
magnitude of 12 (the limit of the telescope-photometer system) were
observed preceding outward from the center as defined by the ID
chart of Hoag, et al (1961). The observed sample accumulated during
the three seasons is practically complete up to the given magnitude.

\section{Data acquisition}

The procedures in both the observations and the reduction were the
same in the three seasons. Each measurement consisted of five
ten-second integrations of each star and one ten-second integration
of the sky for the $uvby$ filters and five ten-second integrations
for the narrow and wide filters, with one ten-second integration of
the sky. Individual uncertainties were also determined by
calculating the standard deviations for each star. The percentual
error in each measurement is, of course, a function of both the
spectral type and the brightness of each star, but they were
observed long enough to secure enough photons to get a S/N ratio
$N/\sqrt{(N)}$ such that the photometric precision is $0.01$ mag in
all cases.

A series of standard stars was also observed on each night to
transform the data into the standard system. The reduction procedure
was done with the numerical packages NABAPHOT (Arellano-Ferro \&
Parrao, 1988) and DAMADAP (Parrao, 2000) which reduce the data into
a standard system, although for the standard bright stars some were
also taken from  The Astronomical Almanac (2006). The chosen system
was that defined by the standard values of Olsen (1983) and the
transformation equations are those defined by Crawford \& Barnes
(1970) and by Crawford \&  Mander (1966). In these equations the
coefficients D, F, H and L are the slope coefficients for $(b-y)$, $
m_1$,  $c_1$ and $\beta$, respectively. B, J and I are the color
term coefficients of $V$, $m_1$, and $c_1$. Those of the 1986 and
1989 seasons have been presented elsewhere (Pe\~na et al. 2006;
Pe\~na and Peniche, 1994, respectively).

Errors of the 2006 season were evaluated by means of the standard
stars observed. These were calculated through the differences in
magnitude and colors, as well as with a linear regression $Y = A + B
* X$ and are presented in Table 1. Emphasis is made that there is a
relatively large discrepancy in V magnitude of HR8086 which was kept
due to its large m1 and c1 values which allow us to evaluate in a
large range the color fits, $V$: (5.5,8.1); $b-y$: (0,0.8); $m_1$:
(-0.1,0.67); $c_1$: (0.07,1.11) and $\beta$: (2.6, 2.9).

\begin{table}[!t]
 \caption{Errors of the 2006 season evaluated by means of the
observed standard stars}
 \begin{center}
  \begin{tabular}{cccccc}
\hline\hline
       &    A      &    B      &    R      &   SD      &  N \\
\hline
$V$    & -0.08251  &  1.01211  &  0.9993   &  0.03781  &  10\\
$b-y$  &  0.00459  &  0.99293  &  0.99709  &  0.01861  &  10\\
$m_1$  &  0.00262  &  0.99155  &  0.99218  &  0.02209  &  10\\
$c_1$  &  0.00488  &  0.99638  &  0.99846  &  0.01906  &  11\\
bt     &  0.11714  &  0.95874  &  0.99131  &  0.01821  &   5\\
\hline \hline
\end{tabular}
\end{center}
\end{table}

Table 2 lists the averaged photometric values of the seventy five
observed stars in the three seasons. Column 1 reports the id of
the stars as listed by WEBDA (Paunzen \&  Mermilliod, 2006),
columns 2 to 5 the Str\"omgren values $(b-y)$, $ m_1$, and $c_1$,
respectively column 6, the $\beta$,  whereas columns 7 to 9 the
unreddened indexes $[m_1]$, $[c_1]$ and $[u-b]$ derived from the
observations. Column 10 lists the spectral types as reported by
WEBDA from several sources and those derived from the Str\"omgren
photometry.

\section{Comparison of the data with the literature values}

\subsection{WEBDA}

A comparison was made with the WEBDA compilation. However, since
basically no previous Str\"omgren photometry had been done on this
cluster, the comparison was made using the existing UBV photometry.
The intersection of both photometric sets was constituted of 74
stars in the V range from 6 to almost 16 magnitude and in the $B-V$
and $U-B$ color indexes from $-0.5$ to $1.6$ and $1.5$ mag,
respectively.  In the $V$ magnitude and $B-V$ vs. $b-y$ diagrams (Figure
1) only three stars (51,54,27) and (91,12,54), respectively, are
openly discordant in each one. There is a slight curvature in the
$\delta$V difference above magnitude 15 which, in our opinion is due
most likely to the measurements of the photographic magnitudes more
than the photoelectric measurements. On the other hand, the $u-b$ vs.
$U-B$ diagram shows a peculiar behavior towards the hotter stars. In
view of the fact that our measurements are the results of three
complete calibrations and given the excellent results among them we
cannot explain this behavior.

\begin{figure}
\begin{center}
\includegraphics[height=6.5cm=width=6.5cm]{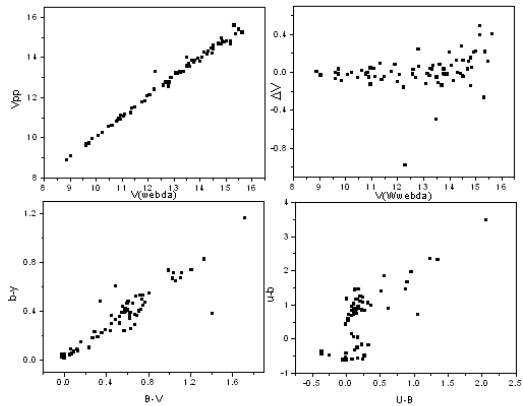}
\caption{Comparison of the $\uv$ photometry with UBV from WEBDA.}
\end{center}
\end{figure}

\subsection{With  Schneider's (1987) $\uv$ photometry}

As was mentioned in the introduction, Schneider (1987) published a
list of 19 measured stars in the $\uv$ system. Hence, a direct
comparison can be made. Figure 2 presents the results of this
comparison; first, the comparison in $V$, $b-y$, $m_1$, $c_1$ and
$\beta$ whereas the last figures the difference between Schneider's
data minus present papers data versus $V$ magnitude from Schneider.
A linear fit to the direct comparison gives the coefficients from
the linear regression $Y = A + B * X $presented in Table 3.

\begin{table}[!t]
 \caption{Linear fit between pp data and Schneider's (1987) $\uv$}
 \begin{center}
  \begin{tabular}{cccccc}
\hline\hline
        &      A     &     B     &   R     &  $\sigma$  &  N\\
\hline
$V$     & -0.08521 & 1.01077 & 0.999 & 0.021 & 6\\
$b-y$   &  0.0088  & 1.0082  & 0.996 & 0.009 & 6\\
$m$     &  0.01455 & 0.90377 & 0.833 & 0.024 & 6\\
$c$     &  0.11518 & 0.8899  & 0.964 & 0.048 & 6\\
$\beta$ & -0.2968  & 1.1027  & 0.981 & 0.011 & 6\\
\hline \hline
\end{tabular}
\end{center}
\end{table}

As can be seen from both Table 3 and Figure 2, the correlation is
adequate. We have not taken into consideration star 5 which they
reported as a miss id. Also, the poorness in the color indexes
could have arisen from the reduced sample Schneider took and, most
likely, the standard stars were taken accordingly. This can be
seen from the small range for $b-y$, $m_1$ and $c_1$ values he
considered. Since we were not restricted to early type stars, we
have observed much larger ranges in the colors and, hence,
obtained more reliable indexes. Nevertheless, despite these
arguments, the differences are within a few hundredths of
magnitude, Table 3.

\begin{figure}
\begin{center}
\includegraphics[height=6.5cm=width=6.5cm]{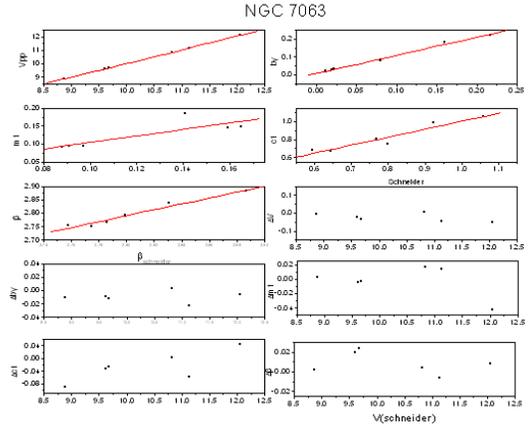}
\caption{Comparison between present paper's data with that of
Schneider (1987).}
\end{center}
\end{figure}

\section{Methology}

As has already been mentioned (Pe\~na et al., 2006), the most
important parameter determined when studying the nature of a cluster
is, beyond a doubt, cluster membership which can be established
using the advantages of Str\"omgren photometry with calibrations
made by Nissen (1988) based on calibrations of Crawford (1975, 1979)
for the A and F stars and of Shobbrook (1984) for early type stars.
These calibrations have been already employed and described in
previous analyses of open clusters (Pe\~na \&  Peniche, 1994). The
determination of physical parameters such as effective temperature,
surface gravity and luminosity has been done in the present study
through the Str\"omgren photometric data reduced to the standard
system, once corrected for interstellar extinction. If the
photometric system is well-defined and calibrated, it will provide
an efficient way to investigate physical conditions. A comparison
with theoretical models, such as those of Lester, Gray \&  Kurucz
(1986, hereinafter LGK86), allows a direct comparison with
intermediate or wide band photometry measured from the stars with
those obtained theoretically for early type stars. LGK86 calculated
grids for stellar atmospheres for G, F, A, B and O stars for the
solar abundance ${\rm [Fe/H]} = 0.00$ in a temperature range from
5500 K up to $50~000$~K. The surface gravities vary approximately
from the Main Sequence to the limit of the radiation pressure in
$0.5$ intervals in $\log g$. They also considered abundances of
$0.1$ solar and $0.001$ solar. A comparison of the photometric
unreddened indexes $(b-y)_0$ and $c_0$ obtained for each star with
such models allows the determination of the effective temperature
$T_e$ and surface gravity $\log g$.

The evaluation of the reddening was done by establishing, as was
stated above, to which spectral class the stars belonged: early (B
and early A) or late (late A and F stars) types; the later class
stars (later than G) were not considered in the analysis since no
reddening determination calibration has yet been developed for MS
stars.  In order to determine the spectral type of each star the
location of the stars in the $[m_1]-[c_1]$ diagram was employed as a
primary criteria. Further analyses were done following the
prescriptions of Lindroos (1980) which merely confirmed our primary
determination. In Table 2 the photometrically determined spectral
class has been indicated. We point out the perfect agreement between
these spectral types and those obtained spectroscopically and
reported by WEBDA.

\section{Results}

The application of the above mentioned numerical packages gave the
results listed in Table 4 in which the ID, the reddening, the
unreddened indexes, the absolute magnitude, the DM and distance, are
listed.  When histograms of the distances are drawn, Figure 3, one
can see that most of the early type stars lie at distance centered
at 760 pc, but with a relatively small spread towards the higher
values. If we consider membership to the cluster to those stars
within one sigma of the mean, we can conclude that most of the
measured stars do belong to the cluster. With respect to the
membership we have determined  practically the same stars that
Schneider (1987) did to be cluster members, although the sample in
consideration is much larger in the present work (75 stars compared
to his 10 stars). Membership is established at the last column of
Table 4. We call attention to the fact that the two late type stars
have been determined as members, W49 and W53, both of F type, are
both metal-poor stars. It should be kept in mind, however, that
these F stars have apparent magnitudes fainter than 13.2 whereas the
cluster members A type stars are one magnitude brighter and the
early type stars belonging to the cluster, even brighter, all
brighter than apparent magnitude 12.0.  At least in the case of W53
it was measured in two seasons, 1986 and 1989 and both seasons give,
independently, large under-abundant metallicities. Hence, although
the star counts were large enough to reach an adequate precision in
all cases their uncertainties are, necessarily larger. More data on
these two stars is needed to settle this apparent paradox.

In order to reach later type stars we should measure fainter stars
which might be done through CCD photometry. In this sense, the
values reported here will serve as secondary standards. At any rate,
the conclusion on distance and age will not change. in going to
fainter stars we will reach the F type stars and solve the puzzle
established by the only two metal-poor stars which lie at the
cluster distance.

\begin{figure}
\begin{center}
\includegraphics[height=6.5cm=width=6.5cm]{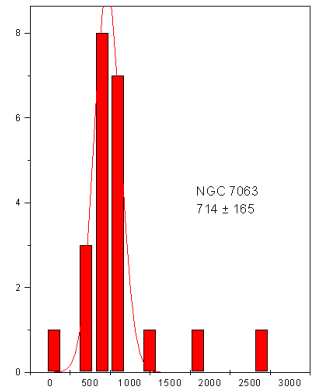}
\caption{Histogram of the distances (X axis, in parsecs) found for
the B and A stars in the direction of NGC 7063. Continuous line is
a Gaussian fit to the data. Its mean value and deviation are
indicated.}
\end{center}
\end{figure}

Once the membership can be established, age is determined after
calculating the effective temperature of the hottest stars.
Temperatures were determined by plotting the location of all stars
on the theoretical grids of LGK86 once we had evaluated the
unreddened colors (Figure 4) for a solar chemical composition. We
have utilized the $(b-y)$ vs. $c_0$ diagrams which allows the
determination of the temperatures with an accuracy of a few
hundredths of degrees. The temperature for the hottest stars, W4,
W47 and W01 are at around 13,000 K ($\log T_e = 4.114$). Hence,
given the calibrations of Meynet, Mermilliod and Maeder (1993) for
open clusters, a log age of 8.146 is found from the relation $-3.6
\log T_e +22.956$ valid in the range $\log  T_e$ between [3.98,
4.25].

\begin{figure}
\begin{center}
\includegraphics[height=6.5cm=width=6.5cm]{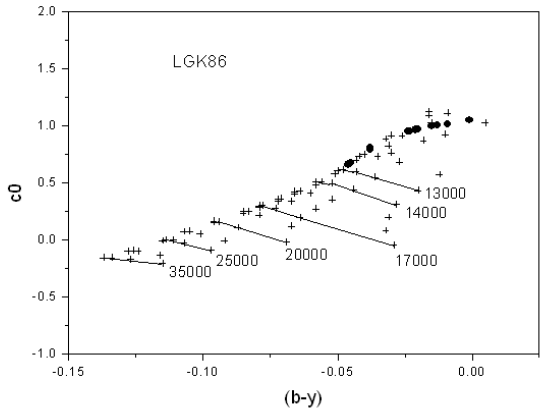}
\caption{Location of the unreddened points (dots) in the LGK86
grids.}
\end{center}
\end{figure}

Figure 5 shows a color-magnitude diagram of the NGC 7063 cluster
considering only the new cluster  members (filled circles)
 and  two theoretical isochrones computed
with a metallicity of [Fe/H]=0.049 (Z=0.02, X=0.73) for two ages of
95 Myr and 125 Myr (continuous lines). The theoretical isochrones
were computed as explained in Fox Machado et al. (2006). In
particular, they were calibrated from [$T_e, \log (L/L_{\odot})$] to
$(B-V, M_{V})$ by using the Schmidt-Kaler (1982) calibration for
magnitudes and the relationship between $T_{\rm eff}$ and $B-V$ of
Sekiguchi \& Fukugita (2000) for the colors.  The obsevational data
in the Str\"omgren photometric system were converted into the
Johnson photometric system by using the transformation relations
given by Turner (1990). Individual star reddenings were used to
obtain the absolute magnitudes of the stars and  an averaged
distance modulus of 9.27 was considered. As can be seen in Fig. 5
the isochrones match  the observed colour-magnitude diagram well.

\begin{figure}
\begin{center}
\includegraphics[height=6.5cm=width=6.5cm]{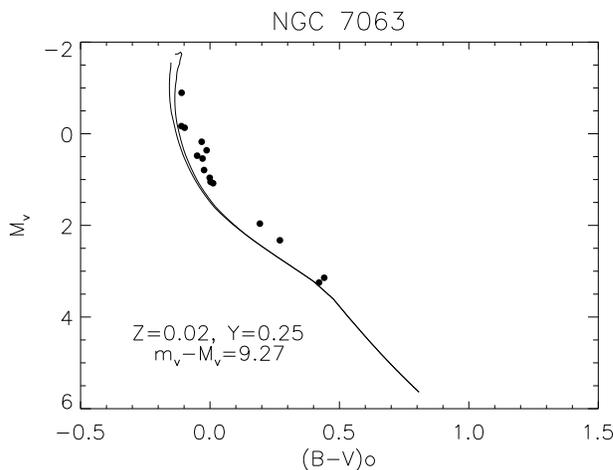}
\caption{Color-magnitude diagram of the NGC 7063 cluster considering
only the new cluster members. The target stars are represented by
filled circles. Two isochrones of 95 Myr (right continuous line) and
125 Myr (left continuous line) computed with  Z=0.02 and Y=0.25 are
shown.  The distance modulus corresponds to that derived in the
present paper. }
\end{center}
\end{figure}

\section{Conclusions}

New $\uv$ photoelectric photometry has been acquired and is
presented for the open cluster NGC 7063. From the 75 observed
stars in the relatively rich field, only a few were determined as
early type stars, either B (15) or A (6). Using the calibration to
determine reddening and distance for these types of stars, a
cluster at 714 pc has been determined. Unreddened indexes in the
LGK86 grids allowed us to determine the effective temperature of
the hottest stars at 13000K.

WEBDA reports the following for NGC 7063: a distance [pc] of 689 (a
distance modulus [mag] $9.47$) and a reddening E(B-V) [mag] of
$0.091$; a $\log$ age $7.977$; on the other hand, Schneider (1987)
reports a distance of $635\pm 30$ pc and a color excess $E(B-V) =
0.08$ mag; these results are quite concordant with our findings: a
mean distance of $722 \pm 105$ pc, which corresponds to a distance
modulus of $9.27$ and a reddening, $E(b-y) = 0.091 \pm 0.039$ mag
which, through the relationship of $E(b-y)= 0.7 * E(B-V)$, yields to
$0.13$ mag, in coarse agreement with the literature. The $\log$ age
we determined, given the youth of the member stars, is of $8.146$.

\acknowledgments

We would like to thank the staff of the OAN for their assistance
in securing the observations. This work was partially supported by
Papiit IN108106 and Conacyt through grants 3925E, E130-3059 and
collaboration programs with the CSIC (Spain) which allowed
reciprocal visits. Typing and proofreading were done by J. Orta,
and J. Miller, respectively. C. Guzm\'an and A. Diaz assisted us
in the computing. This research has made use of the Simbad
databases operated at CDS, Strasbourg, France; NASA ADS Astronomy
Query Form and the WEBDA database, operated at the Institute for
Astronomy of the University of Vienna.

\setcounter{table}{2}
\begin{table*}[!t]\centering
\setlength{\tabnotewidth}{\columnwidth}
  \tablecols{12}
  \setlength{\tabcolsep}{1.0\tabcolsep}
 \caption{$\uv$ photoelectric photometry of NGC 7063}
  \begin{tabular}{rccccccccccc}
    \toprule
 ID & $\langle V\rangle$ & $\langle (b-y)\rangle$ & $\langle m_1\rangle$ & $\langle c_1\rangle$
 & $\langle bt\rangle$ & $[m_1]$ & $[c_1]$ & $[u-b]$ & MK & 207 &  phtm\\
\midrule
1    &  8.882 &  0.022  &  0.094 &  0.684 &  2.756 &  0.101 &  0.680 &  0.882 & B9III & B9IV  & B\\
2    &  9.638 &  0.028  &  0.095 &  0.802 &  2.766 &  0.104 &  0.796 &  1.004 & B8Vn  & AOV   & \\
4    &  9.706 &  0.034  &  0.091 &  0.670 &  2.751 &  0.102 &  0.663 &  0.867 & B8V   & B7V   & B \\
5    & 10.248 &  0.043  &  0.131 &  0.959 &  2.863 &  0.145 &  0.950 &  1.240 & B9.5V & AOV   & B \\
6    & 10.819 &  0.077  &  0.148 &  1.057 &  2.883 &  0.173 &  1.042 &  1.387 & A1V   & A0VNN & B\\
7    & 10.858 &  1.166  &  0.771 &  0.389 &  2.605 &  1.144 &  0.156 &  2.444\\
8    & 10.907 &  0.668  &  0.440 &  0.384 &  2.575 &  0.654 &  0.250 &  1.558 \\
9    & 11.174 &  0.182  &  0.145 &  0.983 &  2.837 &  0.203 &  0.947 &  1.353 &       &       & B \\
10   & 11.774 &  0.355  &  0.141 &  0.421 &  2.641 &  0.255 &  0.350 &  0.859\\
12   & 12.093 &  0.333  &  0.130 &  0.496 &  2.711 &  0.237 &  0.429 &  0.903 &       &       & AF\\
14   & 12.104 &  0.223  &  0.184 &  0.755 &  2.791 &  0.255 &  0.710 &  1.221 &       &       & AF \\
16   & 12.582 &  0.717  &  0.343 &  0.441 &  2.547 &  0.572 &  0.298 &  1.442\\
18   & 13.274 &  0.391  &  0.134 &  0.344 &  2.671 &  0.259 &  0.266 &  0.784\\
19   & 13.691 &  0.712  &  0.302 &  0.570 &  2.626 &  0.530 &  0.428 &  1.487\\
20   & 13.545 &  0.482  &  0.081 &  0.289 &  2.569 &  0.235 &  0.193 &  0.663\\
22   & 13.798 &  0.534  &  0.211 &  0.273 &  2.661 &  0.382 &  0.166 &  0.930\\
25   & 14.149 &  0.735  &  0.231 &  0.096 &  2.438 &  0.466 &  -.051 &  0.881\\
27   & 15.586 &  0.239  &  0.528 &  0.087 &  2.718 &  0.604 &  0.039 &  1.248\\
29   &  9.071 &  0.672  &  0.471 &  0.316 &  2.566 &  0.686 &  0.182 &  1.554 & G9III\\
30   &  9.938 &  0.549  &  0.301 &  0.295 &  2.557 &  0.477 &  0.185 &  1.139 & G8IV\\
31   & 10.114 &  0.068  &  0.145 &  0.977 &  2.858 &  0.167 &  0.963 &  1.297 &  &   & B\\
33   & 10.549 &  0.052  &  0.148 &  1.011 &  2.896 &  0.165 &  1.001 &  1.330 &  &   & B\\
34   & 10.605 &  0.091  &  0.131 &  1.010 &  2.868 &  0.160 &  0.992 &  1.312 &  &   & B \\
35   & 10.922 &  0.187  &  0.172 &  0.955 &  2.866 &  0.232 &  0.918 &  1.381 &  &   & AF\\
36   & 11.097 &  0.091  &  0.179 &  1.032 &  2.874 &  0.208 &  1.014 &  1.430 &  &   & B\\
37   & 11.050 &  0.237  &  0.178 &  0.689 &  2.736 &  0.254 &  0.642 &  1.149 &  &   & AF \\
38   & 11.239 &  0.102  &  0.183 &  0.986 &  2.880 &  0.216 &  0.966 &  1.397 &  &   & B \\
39   & 11.482 &  0.647  &  0.367 &  0.356 &  2.554 &  0.574 &  0.227 &  1.375 &  &   & \\
40   & 11.534 &  0.148  &  0.190 &  0.931 &  2.828 &  0.237 &  0.901 &  1.376 &  &   & AF\\
43   & 12.400 &  0.827  &  0.477 &  0.300 &  2.592 &  0.742 &  0.135 &  1.618 &  &   &  \\
44   & 12.621 &  0.248  &  0.096 &  0.857 &  2.787 &  0.175 &  0.807 &  1.158 &  &   & B \\
45   & 12.759 &  0.741  &  0.519 &  0.167 &  2.527 &  0.756 &  0.019 &  1.531 &  &   &  \\
46   & 12.759 &  0.421  &  0.168 &  0.234 &  2.561 &  0.303 &  0.150 &  0.755 &  &   &  \\
47   & 12.971 &  0.366  &  0.023 &  0.735 &  2.716 &  0.140 &  0.662 &  0.942 &  &   & B \\
48   & 13.208 &  0.500  &  0.209 &  0.445 &  2.600 &  0.369 &  0.345 &  1.083 &  &   &   \\
49   & 13.178 &  0.400  &  0.094 &  0.406 &  2.642 &  0.222 &  0.326 &  0.770 &  &   & AF \\
50   & 13.222 &  0.229  &  0.141 &  0.926 &  2.925 &  0.214 &  0.880 &  1.309 &  &   & B\\
51   & 13.275 &  0.461  &  0.217 &  0.255 &  2.650 &  0.365 &  0.163 &  0.892 &  &   &  \\
52   & 13.306 &  0.410  &  0.155 &  0.426 &  2.637 &  0.286 &  0.344 &  0.916 &  &   &  \\
53   & 13.295 &  0.389  &  0.095 &  0.398 &  2.655 &  0.219 &  0.320 &  0.759 &  &   & AF \\
 \bottomrule
 \end{tabular}
\end{table*}

\setcounter{table}{2}
\begin{table*}[!t]\centering
\setlength{\tabnotewidth}{\columnwidth}
  \tablecols{12}
  \setlength{\tabcolsep}{1.0\tabcolsep}
 \caption{Continued}
  \begin{tabular}{rccccccccccc}
    \toprule
 ID & $\langle V\rangle$ & $\langle (b-y)\rangle$ & $\langle m_1\rangle$ & $\langle c_1\rangle$
 & $\langle bt\rangle$ & $[m_1]$ & $[c_1]$ & $[u-b]$ & MK & 207 &  phtm\\
\midrule
54   & 13.999 &  0.379  &  0.167 &  0.378 &  2.576 &  0.288 &  0.302 &  0.879 &  &   & \\
55   & 13.562 &  0.411  &  0.223 &  0.442 &  2.601 &  0.355 &  0.360 &  1.069 &  &   &  \\
56   & 13.848 &  0.470  &  0.270 &  0.235 &  2.615 &  0.420 &  0.141 &  0.982 &  &   &  \\
59   & 13.757 &  0.395  &  0.125 &  0.365 &  2.571 &  0.251 &  0.286 &  0.789 &  &   &  \\
60   & 13.844 &  0.448  &  0.240 &  0.314 &  2.601 &  0.383 & 0.224 &  0.991 &  &  & \\
62   & 13.949 &  0.378  &  0.219 &  0.370 &  2.798 &  0.340 & 0.294 &  0.974 &  &  &  \\
66   & 14.007 &  0.402  &  0.173 &  0.341 &  2.619 &  0.302 & 0.261 &  0.864 &  &  &  \\
67   & 14.253 &  0.301  &  0.263 &  0.382 &  2.764 &  0.359 & 0.322 &  1.040 &  &  &  \\
71   & 14.315 &  0.528  &  0.175 &  0.296 &  2.707 &  0.344 & 0.190 &  0.878 &  &  &  \\
72   & 14.448 &  0.366  &  0.288 &  0.282 &  2.543 &  0.405 & 0.209 &  1.019 &  &  &   \\
73   & 14.193 &  0.532  &  0.166 &  0.233 &   &  0.336 & 0.127 &  0.799      &  &   &  \\
74   & 14.559 &  0.344  &  0.252 &  0.332 &  2.699 &  0.362 & 0.263 &  0.987 &  &  &  \\
75   & 14.408 &  0.467  &  0.070 &  0.351 &  2.862 &  0.219 & 0.258 &  0.696 &  &  & AF\\
76   & 14.580 &  0.292  &  0.413 &  0.243 &  2.758 &  0.506 & 0.185 &  1.197 &  &  &   \\
81   & 14.676 &  0.352  &  0.225 &  0.270 &  2.698 &  0.338 & 0.200 &  0.875 &  &  &   \\
82   & 14.640 &  0.389  &  0.221 &  0.375 &  2.755 &  0.345 & 0.297 &  0.988 &  &  &  \\
85   & 14.669 &  0.441  &  0.115 &  0.266 &  2.794 &  0.256 & 0.178 &  0.690 &  &  &  \\
86   & 14.975 &  0.259  &  0.399 &  0.145 &  2.661 &  0.482 & 0.093 &  1.057 &  &  &   \\
88   & 14.819 &  0.373  &  0.300 &  0.403 &  2.409 &  0.419 & 0.328 &  1.167 &  &  &   \\
89   & 14.740 &  0.492  &  0.108 &  0.369 &  2.526 &  0.265 & 0.271 &  0.801 &  &  &  \\
91   & 15.234 &  0.479  &  -.043 &  0.338 &  2.506 &  0.110 & 0.242 &  0.463 &  &  & B\\
92   & 14.677 &  0.608  &  0.047 &  0.429 &  2.719 &  0.242 & 0.307 &  0.791 &  &  &   \\
93   & 14.796 &  0.222  &  0.417 &  0.383 &  2.786 &  0.488 & 0.339 &  1.315 &  &  &   \\
99   & 15.154 &  0.472  &  0.520 &  0.017 &  2.737 &  0.671 & -.077 &  1.265\\
101  & 15.390 &  0.300  &  0.459 &  0.077 &  2.721 &  0.555 & 0.017 &  1.127 &  &  & \\
102  & 14.809 &  0.521  &  0.164 &  0.317 &  2.379 &  0.331 & 0.213 &  0.874 &  &  & \\
1086 & 11.818 &  0.474  &  0.175 &  0.440 &  2.657 &  0.327 & 0.345 &  0.999 &  &  &  \\
1127 & 10.943 &  0.316  &  0.168 &  0.452 &  2.671 &  0.269 & 0.389 &  0.927 &  &  &  \\
1127 & 10.955 &  0.306  &  0.151 &  0.442 &  2.667 &  0.249 & 0.381 &  0.879 &  &  & \\
1398 & 13.792 &  0.690  &  0.574 &  0.273 &  2.083 &  0.795 & 0.135 &  1.725 &  &  & \\
1636 & 12.554 &  0.765  &  0.447 &  0.202 &  2.558 &  0.692 & 0.049 &  1.433 &  &  &  \\
j2   & 14.587 &  0.492  &  0.135 &  0.433 &  2.314 &  0.292 & 0.335 &  0.919 &  &  &  \\
 \bottomrule
 \end{tabular}
\end{table*}

\setcounter{table}{4}
\begin{table*}[!t]\centering
\setlength{\tabnotewidth}{\columnwidth}
  \tablecols{12}
  \setlength{\tabcolsep}{1.0\tabcolsep}
 \caption{Reddening and unreddened parameters of NGC 7063}
  \begin{tabular}{rccccccccccc}
    \toprule
ID & $E(b-y)$ & $(b-y)_0$ & $m_0$ & $c_0$ & $\beta$ & $V_0$ & $M_V$ & $DM$ & DST & [Fe/H] & membership\\
\midrule
75 & 0.329 &    0.138 & 0.169 & 0.285 & 2.862 & 12.990 &    7.370 &  5.63 &  134 &         & NM\\
12 & 0.103 &    0.230 & 0.161 & 0.475 & 2.711 & 11.650 &    3.830 &  7.82 &  366 & $-$0.08 & NM\\
37 & 0.031 &    0.206 & 0.187 & 0.683 & 2.736 & 10.920 &    2.610 &  8.31 &  459 &       & NM\\
35 & 0.112 &    0.075 & 0.206 & 0.933 & 2.866 & 10.440 &    1.960 &  8.48 &  496 &       & NM\\
1  & 0.067 & $-$0.045 & 0.114 & 0.671 & 2.756 &  8.590 & $-$0.210 &  8.80 &  575 &       & mbr\\
31 & 0.089 & $-$0.021 & 0.172 & 0.960 & 2.858 &  9.730 &    0.870 &  8.86 &  592 &       & mbr\\
14 & 0.067 &    0.156 & 0.204 & 0.742 & 2.791 & 11.810 &    2.850 &  8.96 &  621 &       & mbr\\
53 & 0.104 &    0.285 & 0.126 & 0.377 & 2.655 & 12.850 &    3.870 &  8.98 &  626 & $-$0.48 & mbr\\
5  & 0.066 & $-$0.023 & 0.151 & 0.946 & 2.863 &  9.960 &    0.930 &  9.03 &  640 &       & mbr\\
33 & 0.065 & $-$0.013 & 0.167 & 0.999 & 2.896 & 10.270 &    1.230 &  9.04 &  644 &       & mbr\\
49 & 0.102 &    0.298 & 0.125 & 0.386 & 2.642 & 12.740 &    3.600 &  9.14 &  674 & $-$0.53 & mbr\\
34 & 0.106 & $-$0.015 & 0.163 & 0.990 & 2.868 & 10.150 &    0.960 &  9.19 &  690 &       & mbr\\
6  & 0.078 & $-$0.001 & 0.171 & 1.042 & 2.883 & 10.480 &    1.070 &  9.41 &  763 &       & mbr\\
2  & 0.066 & $-$0.038 & 0.115 & 0.789 & 2.766 &  9.350 & $-$0.160 &  9.51 &  799 &       & mbr\\
40 & 0.040 &    0.108 & 0.202 & 0.923 & 2.828 & 11.360 &    1.790 &  9.58 &  823 &       & mbr\\
38 & 0.122 & $-$0.020 & 0.220 & 0.963 & 2.880 & 10.710 &    1.090 &  9.62 &  839 &       & mbr\\
4  & 0.080 & $-$0.046 & 0.115 & 0.655 & 2.751 &  9.360 & $-$0.270 &  9.63 &  844 &       & mbr\\
9  & 0.206 & $-$0.024 & 0.207 & 0.944 & 2.837 & 10.290 &    0.650 &  9.64 &  848 &       & mbr\\
36 & 0.100 & $-$0.009 & 0.209 & 1.013 & 2.874 & 10.670 &    1.000 &  9.66 &  857 &       & mbr\\
50 & 0.261 & $-$0.032 & 0.219 & 0.876 & 2.925 & 12.100 &    1.500 & 10.60 & 1316 &       & NM\\
44 & 0.286 & $-$0.038 & 0.182 & 0.803 & 2.787 & 11.390 &    0.130 & 11.26 & 1788 &       & NM\\
47 & 0.412 & $-$0.046 & 0.147 & 0.657 & 2.716 & 11.200 & $-$0.870 & 12.07 & 2592 &       & NM\\
 \bottomrule
 \end{tabular}
\end{table*}

\end{document}